

\def\singlespace{\normalbaselines}
\def\oneandahalfspace{\baselineskip=1.15\normalbaselineskip plus 1pt
\lineskip=2pt\lineskiplimit=1pt}

\def\np{\vfill\eject}
\def\nl{\hfil\break}

\def\nofirstpagenoten{\nopagenumbers\footline={\ifnum\pageno>1\tenrm
\hss\folio\hss\fi}}
\def\nofirstpagenotwelve{\nopagenumbers\footline={\ifnum\pageno>1\twelverm
\hss\folio\hss\fi}}
\def\leaderfill{\leaders\hbox to 1em{\hss.\hss}\hfill}
\def\ft#1#2{{\textstyle{{#1}\over{#2}}}}
\def\frac#1/#2{\leavevmode\kern.1em
\raise.5ex\hbox{\the\scriptfont0 #1}\kern-.1em/\kern-.15em
\lower.25ex\hbox{\the\scriptfont0 #2}}
\def\sfrac#1/#2{\leavevmode\kern.1em
\raise.5ex\hbox{\the\scriptscriptfont0 #1}\kern-.1em/\kern-.15em
\lower.25ex\hbox{\the\scriptscriptfont0 #2}}


\parindent=20pt
\def\narrow{\advance\leftskip by 40pt \advance\rightskip by 40pt}

\def\AB{\bigskip
        \centerline{\bf ABSTRACT}\medskip\narrow}
\def\nonarrower{\advance\leftskip by -40pt\advance\rightskip by -40pt}
\def\AE{\bigskip\nonarrower}

\def\boxit#1{\vbox{\hrule\hbox{\vrule\kern3pt
        \vbox{\kern3pt#1\kern3pt}\kern3pt\vrule}\hrule}}

\def\gtorder{\mathrel{\raise.3ex\hbox{$>$}\mkern-14mu
             \lower0.6ex\hbox{$\sim$}}}
\def\ltorder{\mathrel{\raise.3ex\hbox{$<$}|mkern-14mu
             \lower0.6ex\hbox{\sim$}}}
\def\dalemb#1#2{{\vbox{\hrule height .#2pt
        \hbox{\vrule width.#2pt height#1pt \kern#1pt
                \vrule width.#2pt}
        \hrule height.#2pt}}}

\font\fourteentt=cmtt10 scaled \magstep2
\font\fourteenbf=cmbx12 scaled \magstep1
\font\fourteenrm=cmr12 scaled \magstep1
\font\fourteeni=cmmi12 scaled \magstep1
\font\fourteenss=cmss12 scaled \magstep1
\font\fourteensy=cmsy10 scaled \magstep2
\font\fourteensl=cmsl12 scaled \magstep1
\font\fourteenex=cmex10 scaled \magstep2
\font\fourteenit=cmti12 scaled \magstep1
\font\twelvett=cmtt10 scaled \magstep1 \font\twelvebf=cmbx12
\font\twelverm=cmr12 \font\twelvei=cmmi12
\font\twelvess=cmss12 \font\twelvesy=cmsy10 scaled \magstep1
\font\twelvesl=cmsl12 \font\twelveex=cmex10 scaled \magstep1
\font\twelveit=cmti12
\font\tenss=cmss10
 
 \font\ninebf=cmbx7 scaled \magstep1
\font\ninerm=cmr7 scaled \magstep1 \font\ninei=cmmi7 scaled \magstep1
\font\ninesy=cmsy7 scaled \magstep1 
\font\eightrm=cmr7 scaled 1140 
 
\font\sevenbf=cmbx7 \font\sevenrm=cmr7 \font\seveni=cmmi7
\font\sevensy=cmsy7 

\catcode`@=11
\newskip\ttglue
\newfam\ssfam

\def\fourteenpoint{\def\rm{\fam0\fourteenrm}
\textfont0=\fourteenrm \scriptfont0=\tenrm \scriptscriptfont0=\sevenrm
\textfont1=\fourteeni \scriptfont1=\teni \scriptscriptfont1=\seveni
\textfont2=\fourteensy \scriptfont2=\tensy \scriptscriptfont2=\sevensy
\textfont3=\fourteenex \scriptfont3=\fourteenex \scriptscriptfont3=\fourteenex
\def\it{\fam\itfam\fourteenit} \textfont\itfam=\fourteenit
\def\sl{\fam\slfam\fourteensl} \textfont\slfam=\fourteensl
\def\bf{\fam\bffam\fourteenbf} \textfont\bffam=\fourteenbf
\scriptfont\bffam=\tenbf \scriptscriptfont\bffam=\sevenbf
\def\tt{\fam\ttfam\fourteentt} \textfont\ttfam=\fourteentt
\def\ss{\fam\ssfam\fourteenss} \textfont\ssfam=\fourteenss
\tt \ttglue=.5em plus .25em minus .15em
\normalbaselineskip=16pt
\abovedisplayskip=16pt plus 4pt minus 12pt
\belowdisplayskip=16pt plus 4pt minus 12pt
\abovedisplayshortskip=0pt plus 4pt
\belowdisplayshortskip=9pt plus 4pt minus 6pt
\parskip=5pt plus 1.5pt
\setbox\strutbox=\hbox{\vrule height12pt depth5pt width0pt}
\let\sc=\tenrm
\let\big=\fourteenbig \normalbaselines\rm}
\def\fourteenbig#1{{\hbox{$\left#1\vbox to12pt{}\right.\n@space$}}}

\def\twelvepoint{\def\rm{\fam0\twelverm}
\textfont0=\twelverm \scriptfont0=\ninerm \scriptscriptfont0=\sevenrm
\textfont1=\twelvei \scriptfont1=\ninei \scriptscriptfont1=\seveni
\textfont2=\twelvesy \scriptfont2=\ninesy \scriptscriptfont2=\sevensy
\textfont3=\twelveex \scriptfont3=\twelveex \scriptscriptfont3=\twelveex
\def\it{\fam\itfam\twelveit} \textfont\itfam=\twelveit
\def\sl{\fam\slfam\twelvesl} \textfont\slfam=\twelvesl
\def\bf{\fam\bffam\twelvebf} \textfont\bffam=\twelvebf
\scriptfont\bffam=\ninebf \scriptscriptfont\bffam=\sevenbf
\def\tt{\fam\ttfam\twelvett} \textfont\ttfam=\twelvett
\def\ss{\fam\ssfam\twelvess} \textfont\ssfam=\twelvess
\tt \ttglue=.5em plus .25em minus .15em
\normalbaselineskip=14pt
\abovedisplayskip=14pt plus 3pt minus 10pt
\belowdisplayskip=14pt plus 3pt minus 10pt
\abovedisplayshortskip=0pt plus 3pt
\belowdisplayshortskip=8pt plus 3pt minus 5pt
\parskip=3pt plus 1.5pt
\setbox\strutbox=\hbox{\vrule height10pt depth4pt width0pt}
\let\sc=\ninerm
\let\big=\twelvebig \normalbaselines\rm}
\def\twelvebig#1{{\hbox{$\left#1\vbox to10pt{}\right.\n@space$}}}

\def\tenpoint{\def\rm{\fam0\tenrm}
\textfont0=\tenrm \scriptfont0=\sevenrm \scriptscriptfont0=\fiverm
\textfont1=\teni \scriptfont1=\seveni \scriptscriptfont1=\fivei
\textfont2=\tensy \scriptfont2=\sevensy \scriptscriptfont2=\fivesy
\textfont3=\tenex \scriptfont3=\tenex \scriptscriptfont3=\tenex
\def\it{\fam\itfam\tenit} \textfont\itfam=\tenit
\def\sl{\fam\slfam\tensl} \textfont\slfam=\tensl
\def\bf{\fam\bffam\tenbf} \textfont\bffam=\tenbf
\scriptfont\bffam=\sevenbf \scriptscriptfont\bffam=\fivebf
\def\tt{\fam\ttfam\tentt} \textfont\ttfam=\tentt
\def\ss{\fam\ssfam\tenss} \textfont\ssfam=\tenss
\tt \ttglue=.5em plus .25em minus .15em
\normalbaselineskip=12pt
\abovedisplayskip=12pt plus 3pt minus 9pt
\belowdisplayskip=12pt plus 3pt minus 9pt
\abovedisplayshortskip=0pt plus 3pt
\belowdisplayshortskip=7pt plus 3pt minus 4pt
\parskip=0.0pt plus 1.0pt
\setbox\strutbox=\hbox{\vrule height8.5pt depth3.5pt width0pt}
\let\sc=\eightrm
\let\big=\tenbig \normalbaselines\rm}
\def\tenbig#1{{\hbox{$\left#1\vbox to8.5pt{}\right.\n@space$}}}
\let\rawfootnote=\footnote \def\footnote#1#2{{\rm\parskip=0pt\rawfootnote{#1}
{#2\hfill\vrule height 0pt depth 6pt width 0pt}}}

\def\tenfoot{\tenpoint\hskip-\parindent\hskip-.1cm}

\overfullrule=0pt
\twelvepoint
\def\sbullet{\raise.2em\hbox{$\scriptscriptstyle\bullet$}}
\nofirstpagenotwelve
\hsize=16.5 truecm
\baselineskip 15pt

\def\ft#1#2{{\textstyle{{#1}\over{#2}}}}

\def\a{\alpha_0}

\def\del{\partial}

\def\acrit{\alpha_0^*}

\oneandahalfspace
\rightline{CTP TAMU--10/92}
\rightline{Preprint-KUL-TF-92/8}
\rightline{February 1992}

\vskip 2truecm
\centerline{\bf On Sibling and Exceptional $W$ Strings}
\vskip 1.5truecm
\centerline{H. Lu,$^*$ C.N. Pope,\footnote{$^*$}{\tenfoot Supported in part
by the U.S. Department of Energy, under
grant DE-FG05-91ER40633.} S. Schrans\footnote{$^\diamond$}{\tenfoot
Onderzoeker I.I.K.W.;
On leave of absence from the Instituut voor Theoretische Fysica, \nl
\indent$\,$ K.U. Leuven, Belgium.
}
and X.J.
Wang\footnote{}{\tenfoot }}
\vskip 1.5truecm
\centerline{\it Center
for Theoretical Physics,
Texas A\&M University,}
\centerline{\it College Station, TX 77843--4242, USA.}

\vskip 1.5truecm
\AB\singlespace
We discuss the physical spectrum for $W$ strings based on the algebras
$B_n$, $D_n$, $E_6$, $E_7$ and $E_8$. For a simply-laced $W$ string, we find
a connection with the $(h,h+1)$ unitary Virasoro minimal model, where $h$
is the dual Coxeter number of the underlying Lie algebra. For the $W$ string
based on $B_n$, we  find a connection with the $(2h,2h+2)$ unitary $N=1$
super-Virasoro minimal model.

\AE\oneandahalfspace

\vskip 1.2truecm
\centerline{\tenfoot Available from hepth@xxx/9202060}

\np
\noindent
{\bf 1. Introduction}
\bigskip

     $W$ strings have received a considerable amount of attention recently.
They are described by two-dimensional conformal field theories with non-linear
local symmetry algebras, namely $W$ algebras, which are higher-spin extensions
of the Virasoro algebra.  It has been known for some time that every simple Lie
algebra is the progenitor of an associated $W$ algebra, with the Virasoro
algebra arising from $su(2)$.  Even though simple Lie algebras can be viewed as
``coupled'' $su(2)$ subalgebras, one cannot understand Lie algebras by studying
$su(2)$ alone.   It seems that $W$ strings can similarly be viewed as
``coupled'' Virasoro strings, so, in the same vein, one should not expect the
full richness of string theories to be uncovered by studying the Virasoro
string only.

     The results obtained hitherto have been restricted to the $W_N$  string,
whose symmetry algebra, $W_N=W\!A_{N-1}$, is derived from the Lie algebra
$su(N)\equiv A_{N-1}$ [1,2,3,4,5].  In particular, it has been shown that
the physical spectrum of the $W_N$  string is effectively given by that of
ordinary Virasoro strings, but with a non-standard central charge $c^{\rm
eff}=26-c(N)$, and a set of non-standard values for the spin-2 intercepts
$L_0^{\rm eff}=1-\Delta_{(r,r)}$, where $c(N)=1-6/\big(N(N+1)\big)$ is the
central charge of the $(N,N+1)$ unitary Virasoro minimal model and
$\Delta_{(r,r)}$ are the values of the diagonal entries of the weights
$\Delta_{(r,s)}$ appearing in its Kac table [5].

     In this paper, we generalise these results to $W$ strings whose
underlying $W$ algebras are derived from any of the simply-laced Lie algebras
$A_n\equiv su(n+1)$, $D_n\equiv so(2n)$, $E_6$, $E_7$ and $E_8$.  These string
theories all turn out to be connected with $(h,h+1)$ unitary Virasoro minimal
models, where $h$ is the dual Coxeter number of the corresponding Lie algebra.
In addition we show that for the non-simply-laced Lie  algebras $B_n\equiv
so(2n+1)$, the corresponding $W$ strings are connected  with $(2h,2h+2)$
unitary $N=1$ super-Virasoro minimal models. A similar analysis seems not to be
possible for the $C_n\equiv sp(2n)$, $F_4$ and $G_2$ algebras.

       For any $W$ algebra based on a rank-$n$ Lie algebra, there exists a
realisation in terms of $n$ free scalars.  All the $W$ algebras that we
shall consider have the remarkable property that one of these scalars
appears in the $W$ currents only {\it via} its energy-momentum tensor.
Consequently this energy-momentum tensor may be replaced by an arbitrary
one, $T^{\rm eff}$, with the same central charge.\footnote{$^*$}{\tenfoot In
the case of the $W\!B_n$ algebra considered here, one also needs a free
fermion, in order to realise the algebra.  This, and one of the free
scalars, appear only {\it via}\ their super energy-momentum tensor.}  The
physical-state conditions from the $W$ currents have the effect of
determining the intercept $L_0^{\rm eff}$ of $T^{\rm eff}$, and of
``freezing'' the remaining $(n-1)$ scalars of the realisation of the $W$
algebra. By choosing $T^{\rm eff}$ to be the energy-momentum tensor for an
independent set of free scalars $X^\mu$, one then has a starting point for a
$W$-string theory. These scalars $X^\mu$ acquire the interpretation of being
the coordinates of the target spacetime. The physical spectrum of the $W$
string is thus determined by the values of $L_0^{\rm eff}$, which follow
from the $W$ constraints. Since these values are independent of the explicit
form of $T^{\rm eff}$, it is sufficient for us to choose the simplest
realisation, {\it i.e.}\ to revert to the original $n$-scalar realisation of
the $W$ algebra, in order to calculate them. In the rest of this paper, even
though we shall usually work with the $n$-scalar realisation, it is to be
understood that our results will acquire a string-theoretic interpretation
when the extra $X^\mu$ coordinates are introduced.

\bigskip\bigskip
\noindent{\bf 2. Simply-laced Algebras}
\bigskip
\noindent{\it 2.1 The General Structure of Simply-laced $W$ Strings}
\bigskip

     The discussion of $W$ strings in the case of simply-laced Lie algebras
can be given in a rather general way. In these cases the energy-momentum tensor
for the  corresponding $W$ algebra is given by
$$
T(z)=-\ft12\del{\vec\varphi}\cdot\del\vec\varphi +\a \vec\rho \cdot
\del^2\vec\varphi\ ,\eqno(2.1)
$$
where $\vec\varphi\equiv (\varphi_1,\ \varphi_2,\ldots,\ \varphi_n)$ are  free
scalar fields, and $\vec\rho$ is the Weyl vector ({\it i.e.} half the sum of
the positive roots) of the underlying Lie algebra $g$, which has rank  $n$. The
central charge of this realisation is
$$
c=n+12\, \a^2\,\vec\rho^{\, 2}\ .\eqno(2.2)
$$
Using the Freudenthal strange formula $\vec\rho^{\, 2}=\ft1{12}h\,  {\rm
dim}(g)$, this can be rewritten as
$$
c=n+\a^2\,h\, {\rm dim}(g)\ ,\eqno(2.3)
$$
where $h$ is the dual Coxeter number\footnote{$^\ddagger$}{\tenfoot For
convenience, we recall that the dual Coxeter numbers for the simple Lie
algebras are as follows: $A_n,\ n+1$; $B_n,\ 2n-1$; $C_n,\ n+1$; $D_n,\  2n-2$;
$E_6,\ 12$; $E_7,\ 18$; $E_8,\ 30$; $F_4,\ 9$; and $G_2,\ 4$.} and dim$(g)$ is
the dimension of the Lie algebra $g$, which, for simply-laced algebras, is
given by
$$
{\rm dim}(g)=n(h+1)\ .\eqno(2.4)
$$
In these expressions the standard normalisation, in which the simple roots
have (length)$^2=2$, is being used.

     Anomaly freedom of the $W$-string theory requires that the central  charge
take its critical value, which is determined by the condition that it cancel
the contribution from the ghosts in the BRST quantisation procedure.  For every
current, with spin $s$, generating the $W$ algebra, there is a  contribution
$-2(6s^2-6s+1)$ to the ghostly central charge, implying that the critical
central charge is given by
$$
c^*=2\sum_{\{s\}}\big(6s^2-6s+1\big)\ .\eqno(2.5)
$$
For the $W$ algebras derived from the simply-laced Lie algebras, the set
$\{s\}$ of spins of the generating currents runs over the orders of the
independent Casimir operators [6,7,8]. These are
$$\eqalign{
A_n\ &:\quad 2,\ 3,\ 4,\ \ldots ,\ n+1\cr
D_n\ &:\quad 2,\ 4,\ 6,\ \ldots ,\ 2n-4,\ 2n-2;\ n\cr
E_6\ &:\quad 2,\ 5,\ 6,\ 8,\ 9,\ 12\cr
E_7\ &:\quad 2,\ 6,\ 8,\ 10,\ 12,\ 14,\ 18\cr
E_8\ &:\quad 2,\ 8,\ 12,\ 14,\ 18,\ 20,\ 24,\ 30\ .\cr}\eqno(2.6)
$$
This leads to a general formula for the critical central charge [6]
$$
c^*=2n(2h^2+2h+1)\ .\eqno(2.7)
$$
Using (2.3) and (2.4), this gives the following expression for the
background-charge parameter $\acrit$:
$$
(\acrit)^2={(2h+1)^2\over {h(h+1)}}\ .\eqno(2.8)
$$
{}From now on, we shall always take the critical values for the central charge
and the background charge parameter.

      It was shown in [2,5] that the $W\!A_n$ algebra can be realised in terms
of the currents of the $W\!A_{n-1}$ algebra and one extra free scalar field
(see also [9], where this was first observed for $W\!A_2$, and conjectured for
$W\!A_n$).  We shall show in subsection 2.2 that the $W\!D_n$ algebra may
similarly be realised in  terms of the currents of the $W\!D_{n-1}$ algebra and
one extra free scalar  field, and we shall give evidence in subsection 2.3 for
a similar reduction procedure for the exceptional algebras.  Applying the above
mentioned reduction  recursively, one can then realise these $W$  algebras in
terms of an arbitrary energy-momentum tensor $T^{\rm eff}$ and  $(n-1)$ free
scalar fields. As explained in [1,2,3,5], the $W$ constraints have the effect
of ``freezing'' the $(n-1)$ free scalars.\footnote{$^*$}{\tenfoot Identifying
the frozen coordinates is not always straightforward. Indeed, for a general
choice of basis of the simple roots of the underlying Lie algebra, the unfrozen
coordinate turns out to be a non-trivial linear combination of the scalars
$(\varphi_1,\ldots,\varphi_n)$.} Thus if $T^{\rm eff}$ is  taken to be the
energy-momentum tensor for a new set of free scalar fields  $X^\mu$, one is
left with an effective Virasoro-like string theory, where the $X^\mu$ fields
have the interpretation of being target-spacetime coordinates. This effective
string theory has a non-standard value for the central charge $c^{\rm eff}$,
and the $W$ constraints imply that the spin-2 intercept $L_0^{\rm eff}$ of
$T^{\rm eff}$ takes values from a finite set.  As we shall see, the central
charge $c^{\rm eff}$ is always given by
$$
c^{\rm eff}=1+6(\acrit)^2\ .\eqno(2.9)
$$

      The relation between these $W$ strings and Virasoro minimal models
emerges by substituting the critical value $\acrit$ for the background-charge
parameter given in (2.8) into (2.9), leading to
$$
c^{\rm eff}=26 -\Big[1-{6\over{h(h+1)}}\Big]\ .\eqno(2.10)
$$
Here, 26 is the critical central charge for the Virasoro string and the
remainder is precisely the central charge of the $(h,h+1)$ unitary Virasoro
minimal model. This connection with minimal models is further  strengthened by
the fact that if one rewrites the effective intercepts $L_0^{\rm eff}$ as
$$
L_0^{\rm eff}=1-L_0^{\rm min}\ ,\eqno(2.11)
$$
where 1 is the value for the intercept of the critical Virasoro string,   then
$L_0^{\rm min}$ takes values precisely from a subset of the dimensions of  the
primary fields of the $(h,h+1)$ minimal model. This has been shown in [4,5] for
the $W\!A_n$ strings, and we shall show it below for the $W\!D_n$  strings.

    We shall focus for now on the tachyonic physical spectrum of these
simply-laced $W$ strings. Here ``tachyonic'' signifies states at level 0,
obtained  by acting on the $SL(2,C)$ vacuum with operators of the form
$$
V_{\vec\beta}=e^{\vec\beta\cdot\vec\varphi}\ .\eqno(2.12)
$$
(Recall that since the additional scalar fields $X^\mu$ appear in the $W$
currents only through their energy-momentum tensor, it suffices, when
discussing tachyonic states, to consider just one such extra scalar, which
is in fact the one  that automatically remains in the reduction process that
we described  earlier.) The physical-state conditions for tachyonic states
are given by
$$
\big(W_s\big)_0\big|{\rm phys}\big\rangle=\omega_s\big|{\rm  phys} \big\rangle
\ ,\eqno(2.13)
$$
where $\big(W_s\big)_0$ denotes the zero Laurent mode of the spin-$s$ current
$W_s(z)$, and $\omega_s$ its intercept.  $W_2(z)$ is the energy-momentum
tensor $T(z)$, given in (2.1).

    These intercepts should in principle be determined by requiring that the
nilpotent BRST operator for the algebra annihilate the physical vacuum. In
practice, however, the construction of the BRST operator is very complicated
and has only been performed for the $W_3=W\!A_2$ algebra [10]. The intercepts
can, however, be determined if one knows a particular physical state, since one
can then simply read them off by acting with  $\big(W_s\big)_0$ on that state.
In fact such a physical state has been proposed for the $W\!A_n$ algebra. It is
obtained by acting on the $SL(2,C)$ vacuum with the ``cosmological-constant
operator''
$$
V_{\rm cosmo}=e^{\lambda \acrit \vec\rho\cdot\vec\varphi}\ , \eqno(2.14)
$$
where $\vec\rho$ is the Weyl vector for $A_n$ and $\lambda$ is a certain
constant to be determined. Since one knows the  values of the intercepts for
the $W\!A_2$ algebra, one can explicitly check the existence of such a physical
state in this case. Classical-correspondence considerations presented in [2]
already suggested that such a physical state indeed exists for all the $W\!A_n$
algebras. Stronger evidence towards this assumption was given in [5], where it
was shown that, if the $W\!A_n$-string theory is to be unitary, then this
``cosmological solution'' is necessarily contained in its physical spectrum.
The simplicity of the form of the cosmological operator (2.14) leads us to
conjecture its existence for {\it all} the simply-laced $W$ algebras. We shall
thus assume that such a state is contained in the physical spectrum of the
simply-laced $W$ strings.

    The value of the constant $\lambda$ in the cosmological operator (2.14)
can be computed by an independent argument that enables one to determine  the
spin-2 intercept $\omega_2$ from the structure of the ghost vacuum.  The total
energy-momentum tensor $T^{\rm tot}\equiv T^{\rm mat}+ T^{\rm ghost}$ generates
a linear algebra, and consequently, since the BRST quantisation procedure
requires $T^{\rm tot}$ to annihilate $\big|{\rm phys}\big\rangle \otimes
\big|{\rm vac} \big\rangle_{\rm ghost}$, one can read off the spin-2 intercept
$\omega_2$  as the negative of the intercept for $T^{\rm ghost}$ on the ghost
vacuum, which is defined by
$$
\big|{\rm vac}\big\rangle_{\rm ghost}\equiv \prod_{\{s\}}\, \prod_{m=1}^{s-1}\,
\big(c_s\big)_m \big|0\big\rangle,\eqno(2.15)
$$
where $\big|0\big\rangle$ is the $SL(2,C)$ vacuum and $\big(c_s\big)_m$ are the
Laurent modes of the usual ghost field for the spin-$s$ current. The spin-2
intercept is therefore given by
$$
\omega_2=\ft12 \sum_{\{s\}}\, s(s-1) =\ft16 n h(h+1)\ .\eqno(2.16)
$$
Note that using the Freudenthal strange formula and (2.4), one finds the
important property
$$
\omega_2=2\, \vec\rho^{\, 2}\ . \eqno(2.17)
$$
The value of $\lambda$ can now be determined by requiring that the cosmological
operator (2.14) be a primary field of dimension $\omega_2$ with respect to the
energy-momentum tensor (2.1).  This gives a quadratic equation for $\lambda$.
As will be clear later, the two solutions of this equations are related by a
Weyl reflection, and one can thus, without loss of generality, take one of the
solutions, {\it e.g.}
$$
\lambda={2(h+1)\over 2h+1}\ , \eqno(2.18)
$$
and refer to this solution as the cosmological solution.

     The complete tachyonic physical spectrum of these simply-laced $W$ strings
then follows from this cosmological solution. To see this, one first recalls
[7,8,11] that the eigenvalues $v_s(\vec\beta)$ of a  state  created by any
tachyonic operator (2.12) under the action of $\big(W_s\big)_0$ are invariant
under the action of the Weyl group ${\cal W}$  of the underlying Lie algebra on
the shifted momentum $\vec\gamma$, which is defined  by
$$
\vec\beta=\acrit(\vec\rho+\vec\gamma)\ .\eqno(2.19)
$$
Since the physical-state conditions for tachyons are given by $v_s(\vec\beta)
=\omega_s$, it follows that the action of Weyl group on the shifted momentum
$\vec\gamma$ maps solutions of these physical-state conditions into solutions.
In fact, we know one solution of the tachyonic physical-state conditions, {\it
viz.}\ the cosmological solution; this is a solution by construction. Weyl
reflecting the shifted momentum of this solution,
$$
\vec\gamma^{\rm cosmo}={1\over 2h+1}\ \vec\rho\ , \eqno(2.20)
$$
thus leads to new physical states of the corresponding $W$ string theory. Since
the Weyl vector is not a fixed point of the Weyl group, one can therefore
construct dim$\,({\cal W})$ different tachyonic physical states.  On the other
hand $v_s(\vec\beta)$ is a polynomial of degree $s$ in $\vec\beta$, and thus
also in the shifted momentum $\vec\gamma$, and so it follows that the tachyonic
physical-state conditions will have $\big(\prod_{\{s\}} s\big)$  different
solutions.  Remarkably, this expression, which is the product of the orders of
the independent Casimir operators, is precisely the dimension of the Weyl
group! (Note that this is true for {\it any} simple  Lie algebra.)  We
therefore conclude that the action of the Weyl group on the shifted momentum
(2.19) of the cosmological solution indeed generates the entire tachyonic
physical spectrum.

     Having found the complete tachyonic physical spectrum of these $W$
strings, one can now compute the effective spin-2 intercepts $L_0^{\rm eff}$.
To do this one has to identify the unfrozen coordinate. This requires knowledge
of the explicit form of the realisation of the $W$ algebra.  It is therefore
necessary to continue the discussion of these $W$ strings case by case.

\bigskip
\noindent{\it 2.2 \ $W$ Strings for Classical Simply-laced Algebras}
\bigskip

     Since the $W\!A_n$ string has been treated in detail in [5], we shall just
summarise the results.  In this particular case, the complete physical spectrum
(including {\it all} higher-level states) has been obtained. Physical states
with excitations in the frozen directions have zero norm and thus decouple from
the theory. The remaining physical states have positive semi-definite norm  and
their effective intercepts, at all higher levels, take the same set of
$L_0^{\rm eff}$ values as those for the tachyonic states. Using (2.11), this
leads to the following values of $L_0^{\rm min}$:
$$
L_0^{\rm min}={k^2-1\over4h(h+1)}\ , \qquad k=1,2,\ldots,h-1\ .\eqno(2.21)
$$
These are precisely the diagonal entries $\Delta_{(k,k)}$ of the Kac table of
the $(h,h+1)$ unitary Virasoro minimal model, whose dimensions are
$$
\Delta_{(r,s)}={[(h+1)r-hs]^2-1\over 4h(h+1)}\ ,\eqno(2.22)
$$
with $1\le r\le h-1$ and $1\le s\le h$.

     Let us now turn our attention to the $W\!D_n$ case. The $W\!D_n$ algebra
is generated by currents $W^{(n)}_{2k}(z)$ of spin $s=2k$, with $k=1, \ldots,
n-1$, and  a current $U^{(n)}(z)$ of spin $n$. Since $D_2\, \cong\,  A_1\times
A_1$ is not simple, we shall restrict ourselves to $n\ge 3$. A realisation of
$W\!D_n$ in terms of $n$ free scalar fields $(\varphi_1,\ldots,\varphi_n)$ has
been given by Fateev and Lukyanov in [7,8]. The spin-$n$ current is given by
the Miura-type transformation
$$
U^{(n)}(z)=(\a\del-\del\varphi_n)\,(\a\del-\del\varphi_{n-1})\, \cdots\,
(\a\del-\del\varphi_2)\del\varphi_1\ , \eqno(2.23)
$$
where, as usual, normal ordering is assumed. The $W^{(n)}_{2k}$ currents can
then be read off from the operator-product expansion $U^{(n)}(z)U^{(n)}(w)$:
$$
U^{(n)}(z)U^{(n)}(w)\sim {a_n\over (z-w)^{2n}} +
\sum_{k=1}^{n-1}{a_{n-k} \over (z-w)^{2n-2k} }
\big(W^{(n)}_{2k}(z)+W^{(n)}_{2k}(w)\big)\ .\eqno(2.24)
$$
In this expression $a_k$ are constants, given by $a_k=(-)^{k+1}
\prod_{j=1}^{k-1}[1+2j(2j+1)\a^2]$, which fix the normalisations of the
higher-spin currents to their standard forms.

     Note that the way in which the $W\!D_n$ currents are obtained is rather
different from the procedure that one uses for the $W\!A_n$ algebra.  In that
case, all the currents are obtained directly from a Miura transformation.  In
fact in general for any simple Lie algebra, a classical $W$ algebra can be
obtained from a Miura transformation [12].  In the case of the $W\!A_n$
algebra, one can straightforwardly take the classical currents and interpret
them as currents for the full quantum $W\!A_n$ algebra, by simply replacing
products of fields by normal-ordered products.  A realisation of the classical
$W\!D_n$ can be similarly obtained from a Miura transformation, involving,
however, a pseudo-differential operator [7,8].  In view of this it is simpler
to obtain the currents that generate the full quantum $W\!D_n$ algebra from the
operator-product expansion of $U^{(n)}$ with itself [7,8], according to (2.24).
In our discussion, however, we shall never need to know the explicit form of
the $W^{(n)}_{2k}$ currents for $k\ge2$.  The  only information that we need
about these higher-spin currents is that,  acting on tachyonic operators
(2.12), they give shifted-momentum polynomials  that are invariant under the
action of the Weyl group, as we discussed  earlier.

     From (2.23) and (2.24) it follows that the energy-momentum tensor
$T=W^{(n)}_2$ is given by (2.1), with
$$
\vec\rho=(0,\ 1,\ 2,\ \ldots,\ n-1)\ .\eqno(2.25)
$$
This is indeed the Weyl vector of $D_n$, with simple roots given by
$$
\eqalign{\vec e_i&=\vec\sigma_{n-i}-\vec\sigma_{n+1-i}, \qquad i=1,2,
\ldots,n-1\cr
         \vec e_n&=\vec\sigma_1+\vec\sigma_2\ ,\cr}\eqno(2.26)
$$
with $\{\vec\sigma_i\}$ (${i=1,\ldots,n}$) the canonical orthonormal basis
$\vec\sigma_i\equiv(0,\ldots, 0,1,0,\ldots,0)$ for $R^n$,
where the 1 occurs in the $i$-th entry.

     The reduction procedure mentioned earlier can now easily been proven
since $U^{(n)}$ can be rewritten as
$$
U^{(n)}=\a\del U^{(n-1)}-\del\varphi_n U^{(n-1)}\ ,\eqno(2.27)
$$
where $U^{(n-1)}$ is the spin-$(n-1)$ current of $W\!D_{n-1}$. It then follows
from (2.24) that the $W^{(n)}_{2k}$ currents can, likewise, be written in terms
of the $W\!D_{n-1}$ currents together with one extra scalar field $\varphi_n$.

     Applying this reduction recursively, one can thus express the currents of
the $W\!D_n$ algebra in terms of the currents of $W\!D_3$ together with $(n-3)$
scalar fields $(\varphi_4,\ldots,\varphi_n)$ (recall that we only consider the
case $n\ge 3$). Since $D_3\cong A_3$, the $W\!D_3$ algebra is isomorphic to
$W\!A_3$. We can then use the recursion procedure proven in [2,5] to
realise $W\!A_3$ in terms of an arbitrary effective energy-momentum tensor
$T^{\rm eff}$ together with 2 scalar fields. We can summarise this recursion
procedure schematically as
$$
W\!D_n\rightarrow W\!D_{n-1} \rightarrow \,\cdots\, \rightarrow
W\!D_3 \cong W\!A_3 \rightarrow W\!A_2 \rightarrow W\!A_1= {\rm Virasoro}
\ .\eqno(2.28)
$$
Since the $W\!A_n$ algebras have already been treated, it is convenient to
rewrite the realisation of $W\!D_3$ in the basis in which $W\!A_3$ has been
studied. This can be obtained by the orthogonal transformation
$$
\pmatrix{\vphantom{1\over\sqrt3}\varphi_1\cr\vphantom{1\over\sqrt3}
\varphi_2\cr\vphantom{1\over\sqrt3}\varphi_3\cr} =
\pmatrix{0 & {2 \over \sqrt{6}} & - {1\over \sqrt{3}}\cr
         -{1\over \sqrt{2}} & {1 \over \sqrt{6}} & {1 \over \sqrt{3}} \cr
          {1\over \sqrt{2}} & {1 \over \sqrt{6}} & {1 \over \sqrt{3}} \cr}
\pmatrix{\vphantom{1\over\sqrt3}\widetilde\varphi_1 \cr
\vphantom{1\over\sqrt3}\widetilde\varphi_2 \cr
\vphantom{1\over\sqrt3}\widetilde\varphi_3\cr}
\ .\eqno(2.29)
$$
Here $(\varphi_1,\varphi_2,\varphi_3)$ are the scalar fields of the $W\!D_3$
realisation as defined in (2.23), and $(\widetilde\varphi_1,
\widetilde\varphi_2,  \widetilde\varphi_3)$ are the scalars of the $W\!A_3$
realisation obtained from the Miura transformation as given in [5]. In this
basis the unfrozen coordinate is $\widetilde\varphi_1 = -(\varphi_2 -
\varphi_3)/\sqrt{2}$.  The effective energy-momentum tensor in the case of the
$W\!D_n$ algebra is thus
$$
T^{\rm eff}\equiv -\ft12 \del\widetilde\varphi_1\del\widetilde\varphi_1 +
\ft1{\sqrt2}\acrit \del^2\widetilde\varphi_1 = -\ft14
\big(\del\varphi_2-\del\varphi_3\big)^2 - \ft12\acrit\big(
\del^2\varphi_2-\del^2\varphi_3\big)\ ,\eqno(2.30)
$$
and generates a Virasoro algebra with central charge $c^{\rm eff}$ given by
(2.9). The effective intercept $L_0^{\rm eff}$ for the $W\!D_n$ string is then
given by
$$
L_0^{\rm eff}=-\ft14 ( \beta_2 -\beta_3 )^2-\ft12 \acrit (\beta_2-\beta_3) \
.\eqno(2.31)
$$
Using equations (2.8), (2.11) and (2.19) it follows that
$$
L_0^{\rm min}={\big[(2h+1)(\gamma_2-\gamma_3)\big]^2-1 \over 4h(h+1)}\ ,
\eqno(2.32)
$$
where $\gamma_2$ and $\gamma_3$ are shifted-momentum components of tachyonic
physical states. Note that substituting for the cosmological solution, {\it
i.e.}\ $\gamma^{\rm cosmo}_2=1/(2h+1)$ and $\gamma^{\rm cosmo}_3=2/(2h+1)$,
one finds $L_0^{\rm min}=0$ which is the identity operator of the $(h,h+1)$
Virasoro minimal model.

       As mentioned in the previous subsection, the complete tachyonic physical
spectrum of the $W\!D_n$ string can be obtained from the action of the Weyl
group of $D_n$ on the shifted-momentum $\vec\gamma^{\rm cosmo}$ given in (2.20)
with $\vec\rho$ given by (2.25). The Weyl group of $D_n$ acts on $\vec\rho$ by
permuting its components  and/or changing their signs [13].   The shifted
momenta for all the tachyonic physical states are thus given by
$$
\vec\gamma={1\over 2h+1}\big(0,\ 1,\ 2,\ \ldots,\ n-1\big)\ ,\eqno(2.33)
$$
together with all possible permutations of the components and all possible
combinations of sign changes, making $2^{n-1} n!$ states in all.  Hence writing
$\gamma_2-\gamma_3=k/(2h+1)$, we find that the absolute value of $k$ can take
all integer values in the range $1,\ 2,\ \ldots,\ 2n-3$.  Since  $h=2n-2$ for
$D_n$, it follows that $L_0^{\rm min}$ takes precisely the values given in
(2.21). We thus conclude that the tachyonic physical states of the $W\!D_n$
string are indeed related to the diagonal entries of the Kac table of the
$(h,h+1)$ unitary Virasoro minimal model.

\bigskip
\noindent{\it 2.3 \ $W$ Strings for Exceptional Simply-laced Algebras}
\bigskip

     Since a Miura-type realisation for the $W\!E_6$, $W\!E_7$ and $W\!E_8$
algebras has not yet been constructed, it seems that {\it a priori} the
corresponding string theories cannot at present be discussed. However, the
structure of the $W\!A_n$ and the $W\!D_n$ strings suggests a possible
generalisation to these cases. The main reason why the classical simply-laced
$W$ strings are related to Virasoro minimal models seems to be equation (2.17),
which expresses the spin-2 intercept $\omega_2$ in terms of the Weyl vector of
the underlying Lie algebra. Since this equation also holds for the exceptional
simply-laced $W$ strings, the entire discussion of subsection 2.1 can be
performed for these cases as well.  The only missing link is a recursion
relation which enables one to identify the unfrozen coordinate and prove that
the central charge $c^{\rm eff}$ of the effective energy-momentum tensor
$T^{\rm eff}$ is given by (2.9).

     It is, however, not too difficult to see how such a recursion procedure
might work.  A possible reduction follows from noting that $E_5\cong D_5$,
which leads schematically to
$$
W\!E_8\rightarrow W\!E_7\rightarrow W\!E_6\rightarrow W\!E_5\cong W\!D_5
\rightarrow W\!D_4 \rightarrow W\!D_3\cong W\!A_3 \rightarrow W\!A_2
\rightarrow {\rm Virasoro}\ . \eqno(2.34)
$$
Alternatively, since $E_4\cong A_4$ one can consider
$$
W\!E_8\rightarrow W\!E_7\rightarrow W\!E_6\rightarrow W\!E_5 \rightarrow
W\!E_4\cong W\!A_4 \rightarrow W\!A_3 \rightarrow\cdots
\rightarrow {\rm Virasoro}\ . \eqno(2.35)
$$
In both of these schemes, the reduction is assumed to be proceeding {\it  via}\
the sequence of canonical embeddings $E_8 \supset E_7 \supset E_6  \supset
\cdots$.  (By ``canonical,'' we mean an embedding described by  deleting a
vertex in the Dynkin diagram.)  {\it A priori}, this is not the only route that
the reduction might follow; for example, one has an alternative sequence of
canonical embeddings $E_p \supset A_{p-1} \supset \cdots\supset A_1$. However,
it should be emphasised that one does not have a free choice in deciding which
reduction scheme is to be selected.  Rather, this is dictated by the detailed
form of the Miura transformation.  For example, we have seen in the previous
subsection that the reduction process for $W\!D_n$ singles out $\varphi_n$,
{\it i.e.}\ the field that sits at the uttermost left of the product of factors
in the Miura transformation (2.23), corresponding to the canonical embedding
$D_n \supset D_{n-1}$.  Even though there are other canonical  embedding
sequences, such as $D_n\supset A_{n-1}\supset A_{n-2} \supset  \cdots\supset
A_1$, the Miura transformation imposes the specific reduction scheme (2.28).
Thus for the exceptional $W\!E_p$ algebras, one has to know the detailed form
of the Miura transformation in order to settle the issue of which reduction
scheme is selected.

      No matter what reduction scheme is actually selected by the Miura
transformation, it is natural to expect that  the relation that holds between
$W\!A_n$ and $W\!D_n$ strings and minimal models  should hold also in the  case
of  the exceptional simply-laced $W$ strings. The tachyonic physical spectrum
of the $W\!E_6$, $W\!E_7$ and $W\!E_8$ strings would then be related to the
diagonal primary fields of the $(12,13)$, $(18,19)$ and $(30,31)$ unitary
Virasoro minimal models respectively.

\bigskip
\bigskip
\noindent{\bf 3. Non-simply-laced Algebras }
\bigskip

       The orders of the independent Casimir operators for the
non-simply-laced simple Lie algebras are
$$\eqalign{
B_n\ &:\quad 2,\ 4,\ 6,\ \ldots,\ 2n\cr
C_n\ &:\quad 2,\ 4,\ 6,\ \ldots,\ 2n\cr
F_4\ &:\quad 2,\ 6,\ 8,\ 12\cr
G_2\ &:\quad 2,\ 6\ .\cr}\eqno(3.1)
$$
For $W\!B_n$ a Miura-type realisation has been given in [7]. In addition to
bosonic currents having the spins listed for $B_n$ in (3.1), there is also a
fermionic current with spin $(n+\ft12)$. The first non-trivial example,
$W\!B_2$, was constructed explicitly in [14].

The fermionic spin-$(n+\ft12)$ current $Q^{(n)}(z)$ of $W\!B_n$ plays an
analogous r\^ole to the bosonic spin-$n$ current $U^{(n)}(z)$ in the $W\!D_n$
algebra. It is given by the Miura-type transformation
$$
Q^{(n)}(z)=(\a\del-\del\varphi_n)\,(\a\del-\del\varphi_{n-1})\, \cdots\,
(\a\del-\del\varphi_1)\, \psi\ , \eqno(3.2)
$$
where $\psi$ is a free fermion field. The bosonic currents $W^{(n)}_{2k}$ can
be read off from the operator-product expansion $Q^{(n)}(z)Q^{(n)}(w)$:
$$
Q^{(n)}(z)Q^{(n)}(w)\sim {b_n\over (z-w)^{2n+1}} + \sum_{k=1}^{n} {b_{n-k}
\over (z-w)^{2n+1-2k} }  \big(W^{(n)}_{2k}(z)+W^{(n)}_{2k}(w)\big)\ .\eqno(3.3)
$$
In this expression $b_k$ are constants, given by  $b_k=\prod_{j=1}^{k}
[1+2j(2j-1)\a^2]$. We shall only need to know the  explicit form of the spin-2
current $T(z)=W_2^{(n)}$. For the higher-spin  currents, it suffices for our
purposes to observe that the eigenvalues of  all the currents acting on
tachyonic operators are invariant under the action of the Weyl group of $B_n$.

    The energy-momentum tensor, which can be read from (3.3),
$$
T(z)=-\ft12\del{\vec\varphi}\cdot\del\vec\varphi
+\a\vec\rho\cdot\del^2\vec\varphi+\ft12 \del\psi\,\psi\ ,\eqno(3.4)
$$
has central charge
$$
c=(n+\ft12)\big(1+2n(2n-1)\a^2\big)\ .\eqno(3.5)
$$
In (3.4) $\vec\rho$ is the Weyl vector of the $B_n$ algebra, which takes the
form
$$
\vec\rho=\big(\ft12,\ \ft32,\ \ldots,\ n-\ft12\big)\ ,\eqno(3.6)
$$
with the simple roots being given by
$$\eqalign{
\vec e_i&=\vec\sigma_{n-i}-\vec\sigma_{n+1-i}, \qquad i=1,2,\ldots,n-1\cr
\vec e_n&=\vec\sigma_{1}\ .\cr}\eqno(3.7)
$$
Note that $\vec\rho^{\, 2}=\ft1{12}n(4n^2-1)$, which we have used in writing
(3.5).

     From (3.2), we see that
$$
Q^{(n)}=\a\del Q^{(n-1)}-\del\varphi_n Q^{(n-1)}\ ,\eqno(3.8)
$$
which enables us to express the currents of $W\!B_n$ in terms of those of
$W\!B_{n-1}$ together with the extra scalar $\varphi_n$.  Applying this
procedure recursively, we may therefore realise $W\!B_n$ in terms of the
currents of $W\!B_1$ together with $(n-1)$ scalars $(\varphi_2,
\ldots,\varphi_n)$.  The $W\!B_1$ algebra, which is generated by currents of
spins $\ft32$  and 2, is isomorphic to the $N=1$ super-Virasoro algebra. Thus
replacing these currents, initially realised in terms of the scalar $\varphi_1$
and the fermion $\psi$, by an arbitrary realisation with the same central
charge $c^{\rm eff}=\ft32+3\a^2$, we find that the $W\!B_n$ string is
effectively reduced to $N=1$ worldsheet super-Virasoro strings.
Schematically, this can be summarised by the diagram
$$
W\!B_n\rightarrow W\!B_{n-1} \rightarrow \,\cdots\, \rightarrow
W\!B_2 \rightarrow W\!B_1=\hbox{super-Virasoro}
\ .\eqno(3.9)
$$

     The critical central charge for the $W\!B_n$ string, given by (2.5) with
an appropriate sign change for the contribution from the ghosts for the
spin-$(n+\ft12)$ fermionic current, is
$$
c^*=(2n+1)(8n^2-4n+1)\ .\eqno(3.10)
$$
This leads to the critical value of the background-charge parameter
$$
(\acrit)^2={{(4n-1)^2}\over {2n(2n-1)}}\ .\eqno(3.11)
$$
Using these results, one finds that the central charge of the effective  $N=1$
superstring is
$$
c^{\rm eff}=15-{3 \over 2}\Big[1-{8 \over 4n(4n-2)}\Big]\ .\eqno(3.12)
$$
Here 15 is the critical central charge for the $N=1$ superstring, and the
remainder is the central charge of the $(4n-2,4n)$ unitary $N=1$
super-Virasoro minimal model.  Following the discussion for the simply-laced
case, we shall now compute the effective spin-2 intercepts and rewrite them as
$$
L_0^{\rm eff}=\ft12-L_0^{\rm min}\ ,\eqno(3.13)
$$
where $\ft12$ is the spin-2 intercept for the critical $N=1$ superstring.   We
shall see that $L_0^{\rm min}$ takes values from the diagonal entries of  the
corresponding $(4n-2,4n)$ super minimal model.

     As in the simply-laced case, we shall assume that the cosmological
operator $e^{\lambda\acrit\vec\rho\cdot\vec\varphi}$ creates a physical  state
when acting on the $SL(2,C)$ vacuum.\footnote{$^\ddagger$}{\tenfoot We shall
consider only the Neveu-Schwarz sector here, for which the fermionic current,
since it has half-integer modes, imposes a  physical-state condition that is
automatically satisfied for tachyonic operators.} The constant $\lambda$ can be
determined from the knowledge of the intercept of the spin-2 current. The
latter can be derived by augmenting the ghost-vacuum discussion of subsection
2.1 to include the contribution of the ghosts for fermionic currents. As
explained in [15], the total contribution to the spin-2 intercept for a theory
with both bosonic and fermionic currents is given by
$$
\omega_2=\ft12\!\!\sum_{\{s\}\atop {\rm bosonic}} \!\!\! s(s-1)
-\ft12\!\!  \sum_{\{s\}\atop {\rm fermionic}}\!\!\! (s-\ft12)^2\ .\eqno(3.14)
$$
For the case of the $W\!B_n$ algebra, we therefore find
$$
\omega_2=\ft16 n(4n^2-1)\ .\eqno(3.15)
$$
Solving the resulting quadratic equation for $\lambda$, and without loss of
generality choosing just one of its roots to define the cosmological solution,
we find
$$
\lambda={4n\over 4n-1}\ .\eqno(3.16)
$$

     The physical-state conditions for tachyonic states are invariant under
the action of the Weyl group of $B_n$ on the shifted momentum as defined in
(2.19).  The dimension of the Weyl group is $2^n n!$, which is precisely equal
to the product of the spins of the bosonic currents in the $W\!B_n$ algebra.
Since the physical-state condition from a spin-$s$ current is a polynomial  of
degree $s$ in the shifted momentum, it follows that $2^n n!$ is also the
number of tachyonic physical states.  Thus, since the Weyl vector is not a
fixed point of the Weyl group, we obtain the entire tachyonic spectrum by
Weyl-reflecting the cosmological solution.

     Since the reduction from $W\!B_n$ to $W\!B_1$ freezes the fields
$(\varphi_2, \ldots,\varphi_n)$, it follows that the energy-momentum tensor for
the effective super-Virasoro string theory is given by
$$
T^{\rm eff}=-\ft12\big(\del\varphi_1\big)^2 +\ft12\acrit \del^2\varphi_1
+\ft12\del\psi\, \psi\ .\eqno(3.17)
$$
Thus by acting on tachyonic operators of the form (2.12), we find that
$L_0^{\rm  eff}$ is given by
$$
L_0^{\rm eff}=-\ft12 \beta_1^2 +\ft12 \acrit\beta_1=\ft18(\acrit)^2\big(
1-4\gamma_1^2\big)\ ,\eqno(3.18)
$$
where the second equality is formulated in terms of the shifted-momentum
component $\gamma_1$.  To find the allowed values of $\gamma_1$, one has to
act with the Weyl group on the cosmological solution.  This has the effect  of
permuting the components, and changing their signs in all possible ways  [13],
giving $2^n n!$ possibilities.  Consequently $\gamma_1$ can take  the values
$$
\gamma_1=\pm{2k-1\over 2(4n-1)},\qquad k=1,2,\ldots,n\ .\eqno(3.19)
$$
Using (3.18), we find that $L_0^{\rm min}$, defined by (3.13), takes the values
$$
L_0^{\rm min}={(2k-1)^2-1\over 16n(2n-1)}\ .\eqno(3.20)
$$
The general result for the dimensions $\Delta_{(r,s)}$ of the  primary fields
in the Neveu-Schwarz sector of the $N=1$ super-Virasoro  $(m,m+2)$ minimal
model is
$$
\Delta_{(r,s)}={[(m+2)r-ms]^2-4\over8m(m+2)}\ ,\eqno(3.21)
$$
where $1\le r\le m-1$ and $1\le s\le m+1$, with $(r-s)$ an even integer.
Comparing with (3.20), we conclude that the tachyonic spectrum of the $W\!B_n$
string is related to the $r=s=2k-1$ diagonal entries of the Kac table of the
$(4n-2,4n)$ unitary super-Virasoro minimal model, for $k=1,2,\ldots,n$.
Unlike the simply-laced case, we therefore find only a subset of the diagonal
entries  at this tachyonic level.  It may be that the remaining diagonal
entries  arise from higher-level states, or from the Ramond sector.

     There exists a different $W$ algebra based on $B_n$. It can be obtained
from a Hamiltonian reduction of $B_n$, and is generated by bosonic currents  of
spins $2,4,\ldots, 2n$ only [12]. The spin-2 intercept for the corresponding
string theory can be easily found to be given by $\omega_2 = \ft16
n(n+1)(4n-1)$, which is not equal to $2\vec\rho^{\, 2}$,  where  $\vec\rho$ is
given by (3.6). For the $W\!B_n$ string discussed in detail  above, which has
an additional fermionic spin-$(n+\ft12)$ current, this  important relation
(2.17) {\it is} satisfied.\footnote{$^*$}{\tenfoot Recall  that the
(length)$^2$ of the long simple roots is normalised to 2. } It seems that this
relation between the spin-2 intercept and the Weyl vector is crucial in
establishing the connection between $W$ strings and minimal models.

      In the cases of $C_n$, $F_4$ and $G_2$, equation (2.17) is not satisfied.
One can however, as for $B_n$, add additional currents in the cases of $C_n$
and $F_4$ so that (2.17) is satisfied. For $C_n$, this  can be done by adding a
spin-$(n+1)$ current, but then the resulting field content is precisely that
of the $W\!D_{n+1}$ algebra. For $F_4$, adding a fermionic current of spin
$\ft{17}2$  leads to a spin-2 intercept satisfying (2.17). To show the possible
connection with a minimal model,  however, one needs an explicit realisation of
the corresponding $W$ algebra.

\bigskip\bigskip
\noindent{\bf 4. Conclusion and Discussions}
\bigskip

      In this paper, we have studied some properties of $W$ strings based on
sibling and exceptional simple Lie algebras. We have found that many of the
features encountered previously for $W\!A_n$ strings [5] carry over to these
other examples, notably the connection between the tachyonic spectrum and
minimal models.

      In the case of the $W\!A_n$ string, it was shown that the only
higher-level states that contribute to the physical spectrum are those that
involve excitations exclusively in the unfrozen directions [5]. Moreover  these
higher-level physical states have a set of effective  spin-2 intercepts
identical to the corresponding set for the tachyonic level. Consequently the
physical states at level $\ell$ have (mass)$^2$ given by $2\ell$ plus the
(mass)$^2$ of the corresponding tachyonic physical states. However, this
notion of mass is rather problematical, owing to the presence of background
charges; see refs.\ [3,5] for further discussion of this point. In view of our
findings  it is natural to expect that higher-level physical states will again
arise only in the unfrozen directions for the $W$ strings discussed in this
paper. Strong evidence for the unitarity of $W\!A_n$ strings was presented  in
[5]; it is likely that the $W$ strings considered here are unitary also.

\bigskip\bigskip
\centerline{\bf ACKNOWLEDGMENTS}
\bigskip

      We thank HoSeong La for his careful reading of the manuscript.
Stany Schrans is very grateful to the Center for Theoretical Physics,
Texas A\&M University, for hospitality, and to the Belgian National Fund for
Scientific Research for a travel grant.

\np
\singlespace
\centerline{\bf REFERENCES}
\frenchspacing
\bigskip

\item{[1]}C.N.\ Pope, L.J.\ Romans and K.S.\ Stelle, {\sl Phys.\
Lett.}\ {\bf 268B} (1991) 167;\nl
{\sl Phys.\ Lett.}\ {\bf 269B} (1991) 287.

\item{[2]}S.R.\ Das, A.\ Dhar and S.K.\ Rama, {\sl Mod.\ Phys.\ Lett.}\
{\bf A6} (1991) 3055;\nl
``Physical states and scaling properties of $W$  gravities and $W$ strings,''
TIFR/TH/91-20.

\item{[3]}C.N.\ Pope, L.J.\ Romans, E.\ Sezgin and K.S.\ Stelle, ``The $W_3$
String Spectrum,'' preprint CTP TAMU-68/91, to appear in {\sl Phys.\ Lett.\ }
{\bf B}.

\item{[4]}S.K.\ Rama, {\sl Mod.\ Phys.\ Lett.}\ {\bf A6} (1991) 3531.

\item{[5]}H.\ Lu, C.N.\ Pope, S.\ Schrans and K.W.\ Xu, ``The Complete Spectrum
of the $W_N$ String,''  preprint CTP TAMU-5/92, KUL-TF-92/1.

\item{[6]}F.A.\ Bais, P.\ Bouwknegt, M.\ Surridge and K.\ Schoutens, {\sl
Nucl.\
Phys.}\ {\bf B304} (1988) 348;\nl
{\sl Nucl. \ Phys.}\ {\bf B304} (1988) 371.

\item{[7]}S.L.\ Lukyanov and V.A.\ Fateev, {\sl Sov.\ Scient.\ Rev.}\ {\bf
A15} (1990).

\item{[8]}S.L.\ Lukyanov and V.A.\ Fateev, {\sl Sov.\ J.\ Nucl.\ Phys.}\ {\bf
49} (1989) 925.

\item{[9]}L.J.\  Romans, {\sl Nucl.\  Phys.}\ {\bf B352} (1991) 829.

\item{[10]}J.\ Thierry-Mieg, {\sl Phys.\ Lett.}\  {\bf 197B} (1987) 368.

\item{[11]}V.A.\ Fateev and S.\ Lukyanov,  {\sl Int.\ J.\ Mod.\  Phys.}\ {\bf
A3} (1988) 507.

\item{[12]}V.G.\ Drinfeld and V.V.\ Sokolov, {\sl Journ.\ Sov.\ Math.}\ {\bf
30}
(1985) 1975.

\item{[13]}J.E.\ Humphreys, ``Introduction to Lie Algebras and Representation
Theory'', Springer-Verlag (1987).

\item{[14]}J.M.\ Figueroa-O'Farrill, S. Schrans and K. Thielemans, {\sl Phys.\
Lett.}\ {\bf 263} (1991) 378.

\item{[15]}H.\ Lu, C.N.\ Pope, X.J.\ Wang and K.W.\ Xu, ``Anomaly Freedom and
Realisations for Super-$W_3$ Strings,''  preprint CTP TAMU-85/91.

\bye